# Evolutionary and topological properties of gene modules and driver mutations in a leukemia gene regulatory network

Anthony Szedlak[a], Nicholas Smith[b], Li Liu[c], Giovanni Paternostro[d,1], Carlo Piermarocchi[a,1]

[a]Dept. of Physics and Astronomy, Michigan State University; [b]Salgomed Inc., Del Mar, CA; [c]College of Health Solutions, Arizona State University; [d]Sanford Burnham Prebys Medical Discovery Institute, La Jolla, CA

## Abstract

The diverse, specialized genes in today's lifeforms evolved from a common core of ancient, elementary genes. However, these genes did not evolve individually: gene expression is controlled by a complex network of interactions, and alterations in one gene may drive reciprocal changes in its proteins' binding partners. We show that the topology of a leukemia gene regulatory network is strongly coupled with evolutionary properties. Slowly-evolving ("cold"), old genes tend to interact with each other, as do rapidly-evolving ("hot"), young genes, causing genes to evolve in clusters. We argue that gene duplication placed old, cold genes at the center of the network, and young, hot genes on the periphery, and demonstrate this with single-node centrality measures and two new measures of efficiency. Integrating centrality measures with evolutionary information, we define a medically-relevant "cancer network core," strongly enriched for common cancer mutations ($p = 2 \times 10^{-14}$). This could aid in identifying driver mutations and therapeutic targets.

Complex networks | Evolution | Genetics | Cancer | Leukemia

## Significance

We found strong relationships between the topological and evolutionary properties of an acute myeloid leukemia (AML) gene regulatory network (GRN). Interacting genes tend to have similar evolutionary ages and rates, causing the GRN to segregate into slowly-evolving ("cold"), old gene communities and rapidly-evolving ("hot"), young gene communities. The coldest, oldest communities are centrally-located and enriched for fundamental functional groups, whereas the hottest, youngest communities are peripheral and enriched for tissue-specific functions. We also found that driver mutations for AML in particular, and for cancer in general, tend to be in the cold, central "cancer core." The identification of this core could aid in targeted therapies for AML.

## Additional Information

Author contributions: G.P. and C.P. conceptualized and designed the study; L.L. calculated all evolutionary rates; A.S. and N.S. conducted all remaining calculations and analyzed the data; A.S., N.S., L.L., G.P. and C.P. drafted and revised the manuscript.

N.S. is an employee of Salgomed Inc., and C.P. and G.P. own equity in Salgomed Inc. The remaining coauthors have no competing financial interests.

[1]To whom correspondence may be addressed. Email: carlo@pa.msu.edu or giovanni@sbpdiscovery.org.

**Introduction**

The evolutionary history of a gene can be mapped in various ways. The absolute evolutionary rate, for example, can be computed from observed differences in orthologs across species in the context of their phylogenetic relationships (1). The age of a gene can be measured by tracing when the gene first appeared in the phylogenetic tree (2). These quantities allow researchers to chronicle the journey of individual genes across evolutionary history.

But genes do not exist, and therefore do not evolve, in isolation. Mutations in a transcription factor may affect the expression of the genes it regulates, since changes in a protein's amino acid sequence can cause it to lose compatibility with former binding partners, and gain compatibility with others. Accumulation of these alterations can lead to changes in fitness and, eventually, speciation. The evolution of individual genes is thus coupled with the evolution of the structure of the organism's gene regulatory network (GRN), and network properties should be related to the evolutionary properties of its constituent nodes and edges.

It has been proposed that GRNs grow and evolve incrementally via gene duplication followed by mutation and functional divergence (3-7), although changes may have occasionally arrived in bursts, as in whole-genome duplication (8). This time-dependent network formation suggests that GRNs are composed of a core of ancient, conserved genes with fundamental functions, and younger, peripheral genes with species- or cell type-specific function, which mutate frequently until the functions of the newly created pathways are optimized. These mutations can cause changes in GRNs by creating, removing, reassigning, or changing other properties of nodes and edges.

Fraser et al. demonstrated that interacting pairs of proteins have similar evolutionary rates (9). This constraint is likely driven by the necessity of coevolution, since a change in one protein's sequence may require a corresponding change in its partner's sequence in order for the pair to remain compatible. Daub et al. showed that genes which are part of many biological pathways have lower evolutionary rates than genes which belong to few or no known pathways, further supporting the idea that related genes share similar evolutionary properties (10). It has also been shown that evolutionary rates are weakly, but significantly, negatively correlated with degree, closeness centrality, and betweenness centrality (network measures which quantify the importance of individual nodes in different ways) (11, 12), and that essential genes have high centrality and low evolutionary rates (13).

It has been noted that the development of cancer cells within an individual is also a type of evolution, based on genetic variation and natural selection, but with many fundamental differences from the evolution of species, primarily the shorter time scale, the absence of sexual recombination, and the cells' atavistic nature. In other words, cancer represents a partial return to a unicellular state, the only form of life until about 600 million years ago (2, 14-16). The first two differences limit the scope of cancer cell evolution, while the third suggests that modifying existing genes or sets of genes might be sufficient to revert to a unicellular-like state (2, 16).

Armed with an increasingly comprehensive knowledge of the sets of genes most frequently mutated in various cancers (17), we can begin to study and understand their network properties. It was previously shown, for example, that genes which are commonly mutated in acute myeloid leukemia (AML) are located significantly closer to each other in their GRN than average (18). However, a general understanding of how network topology relates to both types of evolution remains elusive.

Here, our goal is to establish quantitative relationships between the evolutionary history of genes and their topological properties in an AML-specific GRN, AML 2.3 (18), both in organismal evolution (using evolutionary rates and ages) and in cancer cell evolution (using sets of common driver mutations). The network properties of the genes changing in the two types of evolution are compared. We demonstrate that evolutionary rates and ages of genes are not uniformly distributed across the network, but are naturally organized in communities with well-defined evolutionary characteristics: old genes cluster with old genes, and young cluster with young. Likewise, "cold genes" (genes with low evolutionary rates) cluster with cold genes, and "hot genes" (genes with high evolutionary rates) cluster with hot genes. This segregation also exists for functional groups found to be enriched by the online gene ontology tool DAVID (19). In terms of network topology, we show that genes and functional groups which are old and cold tend to be central, and those which are young and hot tend to be peripheral. We demonstrate this with traditional single-node centrality measures, and find that PageRank (20), a finite-range centrality measure, shows stronger biological significance than degree (a local measure) and betweenness centrality (a global measure). We also introduce two new network measures, the set efficiency (18) and the interset efficiency, which quantify the mean distance between nodes within a single set and between two sets, respectively. Finally, we use these relationships to examine the evolutionary and network properties of common cancer mutations, allowing us to identify a "cancer network core" which is significantly enriched for recurrent mutations.

**Results**

While computed differently, a gene's evolutionary rate (ER) and its age are related. Young genes and functional groups with novel functions need time to fine-tune their properties in order to optimize the fitness of the host organism, so young genes tend to be hot. Likewise, old genes and functional groups with fundamental roles, such as protein translation, have had enough time to sufficiently optimize their functions, and so should change very slowly. As expected, the ERs and ages of the genes present in AML 2.3 are strongly correlated ($R = 0.504, p < 10^{-300}$).

As previously demonstrated (9), interacting genes tend to have similar ERs and ages. The distributions of differences in ERs and ages between genes linked by an edge in AML 2.3 are significantly closer to zero than those of degree-preserving randomizations of the same network, with an approximate z-score of –96.8 for differences in ER and –72.0 for differences in age (see Figs. S1-S4). This tendency for connected genes to have similar ERs and ages hints that there may be large-scale segregation between clusters of old, cold nodes and young, hot nodes. Indeed, this is reflected in the natural community structure present in AML 2.3, as well as in biological functional groups present within these communities. Fig. 1A shows that the ER distribution of translational elongation genes is noticeably left-shifted relative to the ERs of all genes, indicating that it hosts slowly evolving. Transmembrane genes are much younger than average, as shown in Fig. 1B.

Table 1 lists the ER and age properties for the ten largest communities in AML 2.3, and for the three most significantly enriched functional groups found within each community. Fig. 1C summarizes the main results of Table 1. The ERs and ages for many of these functional groups reflect their biological functions. Zinc finger proteins, for example, are involved in a large number of heterogeneous cellular processes (21), so their genes need to adapt more often than genes with very specific singular functions.

They also have a particularly high rate of duplication and loss, so while the family itself is old (found in animals, plants (22), and fungi (23)), individual genes in this family are young (24). Genes involved in transcriptional regulation must also be flexible enough to tune the expression of target genes in response to environmental changes over time (25, 26). Conversely, the most fundamental functional groups have experienced few changes since early single-celled lifeforms. Functional groups such as mRNA metabolic process (27) and translational elongation (28) are old and stable, having long ago optimized their functions.

The same analysis from Table 1 was conducted for a normal hematopoietic stem cell network (see Table S1B and Methods). This normal network is of lower quality than AML 2.3 because of the reduced number of samples; however, it serves as a qualitative control and further validates the results of our analysis. Table S2 compares Tables S1A and S1B between the AML and normal networks. Several of the same functional groups are enriched in both networks, and each has two enriched blood-specific functional groups (lymphocyte activation and hemoglobin complex for AML, and regulation of leukocyte activation and platelet alpha granule for normal). The lower quality of the normal network is evident in the p-values, as the findings for AML 2.3 are more significant.

Traditional single-node centrality measures such as degree, betweenness centrality (29), and PageRank (20) show small but significant correlation with ERs and ages, with the oldest, coldest genes being the most central (see Tables S3A and S3B). Grouping genes by functional group leads to stronger relationships, the clearest of which is between the mean PageRank and mean age, shown in Fig. 2A (Pearson's $R = -0.75$, $p = 1 \times 10^{-5}$; Spearman's $\rho = -0.86$, $p = 5 \times 10^{-8}$). The corresponding values for degree and betweenness centrality are not significant. The plots of each centrality measure (degree, betweenness centrality, and PageRank, as well as the average of all three measures) versus ER and age are shown in Figs. S9-S16. These three centrality measures are related, but differ in their global reach. Degree is completely local, only dependent on the number of neighbors of a gene; betweenness centrality is global, requiring information from the entire network; but PageRank is between these extremes, influenced by all genes but with more weight granted to those genes which are near-by. The strong correlation between PageRank and ER thus may be explained by the presence of communities in the GRN, since community structure is strongly correlated with ER, as shown in Table 1.

Because of the strong correlation between a gene's history and that of its neighbors, genes are expected to evolve in groups rather than as individuals, which should be evident in the network. The *set efficiency*, the mean of the inverse distance between all pairs of nodes in a set (see Methods), is shown in Fig. 2B for genes ranked from coldest to hottest, and Fig. S6 for genes ranked from oldest to youngest. This indicates that the oldest, coldest genes tend to be close, separated by approximately four directed edges, significantly smaller than the network average of approximately six. The set efficiency monotonically declines as hotter, younger genes are included.

Furthermore, the oldest functional groups efficiently exchange information with each other, and the youngest functional groups are distant from the oldest functional groups as well as from each other. Fig. 2C shows the *interset efficiency*, the mean of the inverse distance from all nodes in one set to all nodes in another (see Methods), between all pairs of functional groups from Table 1, where the functional groups are sorted from oldest to youngest. (Note that the diagonal terms of the interset efficiency matrix are the set efficiencies of the functional groups.) The abundance of connections between old functional groups and the scarcity between old-and-young and young-and-young functional groups suggests that during the course of human evolution, the primitive gene regulatory network began as a core of fundamental genes and pathways. As genes duplicated and mutated, novel functions arose and eventually, through selective duplications, deletions, mutations, and rewirings, novel regulatory pathways emerged, growing outward from these ancient genes. This would place the oldest genes near the middle of the network and the youngest genes toward the periphery.

These observations can also be used to understand the action of disease-causing genes. Diseases can be caused by mutations in very young and hot functional groups, for example the hemoglobin complex in the top right corner of Fig. 1C, which is the functional group to which the inherited hemoglobinopathies belong.

The evolutionary and network properties of cancer-causing mutations are, however, very different. As mentioned, cancer cells undergo a distinct type of evolution, that of the cells within an individual patient (14, 15). In Fig. 3A, we show that common AML-related mutations (30) are enriched in the top 5% of genes when ranked from highest to lowest average centrality (see Methods), identifying five out of 22 mutations with $p = 4 \times 10^{-3}$. Ranking genes from lowest to highest evolutionary rate also performs well, capturing seven out of 22 mutations in the top 5% with $p = 7 \times 10^{-5}$. But ranking genes by combining these measures in a composite score, computed from the mean of the average centrality and inverse evolutionary rate, identifies nine out of 22 mutations in the top 5% with $p = 5 \times 10^{-7}$. Furthermore, this behavior is also observed when we consider a larger list of 198 genes relevant to 21 different cancer types (17), as shown in Fig. 3B. The corresponding number of mutations are 36 using average centrality ($p = 1 \times 10^{-11}$), 30 using evolution rate ($p = 5 \times 10^{-8}$) and 40 using the composite score ($p = 2 \times 10^{-14}$).

We are therefore capturing general properties of cancer causing genes. The composite score can be used to obtain an evolutionarily conserved and central core of the network, where the enrichment for cancer-causing mutations is strongly significant. Using the composite ranking we define the top 5% partition, which is the most enriched for cancer mutations, as the "cancer network core." The properties we describe for the core do not depend on the precise definition of this partition; significant enrichments for cancer mutations are also present if we analyze the top 10% of the distribution. The interactions between the cancer network core and other sets of genes of equal size corresponding to the composite score ranges of Panels 3A and 3B are shown in Panel 3C. This is a representation of the AML Network in which the 20 partitions in Figs. 3A and 3B are represented by single nodes, and the width of each edge is proportional to the sum of all edge weights between the partitions. The edges were sorted by their weights and added to the network until each node had at least one edge. The central node in Fig. 3C, representing the cancer network core, is also the most connected node and its edges have the largest weights. The set efficiency and average interset efficiencies are also significantly higher in this cancer core compared to the other partitions, confirming the centrality of the core.

The core genes are significantly enriched in some functional groups, being over-represented in colder, older functional groups and under-represented in younger, hotter functional groups, especially membrane proteins (Fig. 3D). This may be explained by the fact that membrane proteins have underlying differences in their ERs that makes using the composite score (weighted to find colder genes) ineffective. It is therefore clear that the definition of the cancer network core could be eventually systematically improved using functional information. This may be possible in the future when more data is made available, allowing us to construct a larger set of comparable networks for different cancer types.

## Discussion

We have shown that slowly evolving, old genes tend to interact with each other, and frequently evolving, young genes tend to interact with each other. While these findings are mainly derived from an analysis of the AML network, they were broadly confirmed in a normal hematopoietic network and are consistent with previous reports (13). This supports the hypothesis that GRNs evolved and grew via gene duplication, in which a core of ancient genes with fundamental functions (such as cell cycle regulation and management of RNA) spawned new sets of genes. As these new genes' functions became more refined and specialized, new pathways emerged. Many rounds of this process would place the ancient genes in the center of the new network, and the newer elements would tend to be on the network's periphery.

Our data also show that genes which are conserved in organismal evolution are more likely to be frequently mutated in cancer cells that evolve within a single organism, which points to the high degree of adaptation to multicellular life of the core of the cellular network that cancer is subverting. The result is consistent with the view that cancer is a reversion to primitive unicellular state by modifications of older groups of genes.

No gene is an island. A real understanding of the evolution of a genome only comes from studying its constituent genes in the context of the underlying complex network of interactions rather than as independent entities. This suggests that cancer should be considered a disease of the cancer network core, rather than of individual genes, and that therapies should aim to preferentially affect the activity of the core, rather than individual targets. This is a testable hypothesis because suitable drugs and methods to design drug combinations are increasingly available (31).

Although this analysis focused on an AML-specific network, several results, including the enrichment of genes recurrently mutated in 21 cancer types, show that our findings are of general relevance. As network reconstruction methods continue to improve and more high quality networks become available, we expect to find more evidence of how evolution shapes the topology of gene regulatory networks.

## Conclusions

We demonstrated that the evolutionary history of human genes had a significant impact on the development of GRNs. We provided strong evidence for the coevolution of interacting genes, and showed that this constraint produces natural community structures with well-defined evolutionary characteristics. These characteristics are also present in the functional groups enriched within these communities. The enriched functional groups exchange information with each other with differing efficiency: the oldest, coldest functional groups are close to each other, whereas the youngest, hottest functional groups are more dispersed.

We also introduced a composite measure, which integrates single-gene evolutionary information with network centrality. This measure was used to identify a cancer network core which shows significant enrichment of common cancer mutations in the oldest, coldest, most central genes, i.e. genes with precisely the opposite properties of those which mutate most frequently and drive the evolution of multicellular organisms. Integrating evolutionary and network centrality information could therefore help in determining whether newly discovered mutations contribute to carcinogenesis (so called driver mutations) or are simply passengers. When drugs directly affecting the mutations are not available, we could prioritize interventions acting on the genes in the cancer core.

**Methods**

**Evolutionary rate and age.** To compute the evolutionary rate (ER) of a gene, we first calculated the absolute ER for each amino acid position of the protein it encodes using the method from Kumar et al. (1). Given the multiple alignment at an amino acid position in 46 species (32), its ER equals the number of different residues divided by the total evolutionary time span, based on a known phylogenetic tree (1). The ER of a gene is the average of ERs over all amino acid positions, in units of the number of substitutions per amino acid site per billion years. The ER value ranges from ~0.011 (most conserved) for LSM23 to ~6.928 (least conserved) for CDRT15. Ages, taken from Chen et al. (2), were estimated from comparing the human genome to the genomes of 13 major clades with origins at different points along the human clade, indexed 0 (oldest) through 12 (youngest). A gene's age was determined by searching for the earliest time at which an orthologous gene appears in an organism which branched from the human clade.

**Network and DAVID functional groups.** The gene regulatory network used in this analysis, "AML 2.3", is a partially directed, weighted acute myeloid leukemia (AML) GRN (18). This network was chosen primarily for its quality. It was constructed from more than 1,800 patients across 12 studies from both microarray and RNA-seq gene expression measurements in AML cells, and it is a high confidence network. Additionally, a network constructed from data from a single cell type (in this case, myeloblast cells) focuses on the most active, relevant interactions in that cell type.

A weighted, directed, modularity-based community-finding algorithm was used to divide the 10,062 genes into 133 communities (33). A spy plot of the adjacency matrix after community sorting is shown in Fig. S5. The ten largest communities were selected for further analysis (see Table 1). The individual communities were then provided to the DAVID functional annotation tool to identify enriched functional groups in the communities (19). The top three distinct enriched functional groups with Benjamini values less than $10^{-4}$ in each community are also included in Table 1.

Communities and functional groups in Table 1 labeled "cold" and "hot" have significantly lower and higher evolutionary rates (ERs) than the network average, respectively. Likewise, groups of genes labeled "old" and "young" are significantly older and younger than the network's average age, respectively. A one-tailed significance level of $p < 10^{-3}$ in the difference from the mean was chosen for both ER and age. The Kolmogorov-Smirnov (KS) statistic and p-value were also computed for each community and functional group to quantify the difference between the distribution of all genes and the distribution of each set of genes, which is shown Table S1A. Some example functional group distributions are shown in Fig. 1A and 1B, and all distributions are shown in Figs. S7 and S8 for ERs and ages, respectively. A summary of the ERs and ages of the enriched functional groups in Table 1 is shown in Fig. 1C. The same analysis was conducted for normal hematopoietic stem cell network built from five datasets (GSE48846, 2666, 33223, 24759, and 30376) using the same method as for AML 2.3, with the data reported in Table S1B.

**ERs and ages between interacting genes.** To determine the significance of the correlation between ERs and gene-gene interactions, the difference in evolutionary rates between all gene pairs connected by an edge was computed for AML 2.3 as well as for degree-preserving randomizations of AML 2.3. Fig. S1 shows the distribution of $(ER_j - ER_i)$ for all gene pairs $(i,j)$ which are connected by an edge $j \to i$ in AML 2.3 (green distribution), as well as for all pairs of genes in one degree-preserving randomization of the same network (purple distribution). Note that the distributions are asymmetric because AML 2.3 is a directed network. The real distribution of ER differences has a smaller standard deviation than for the randomized network, meaning that difference in evolutionary rates between interacting genes is small on average, in agreement with Fraser et al. (9). To quantify the significance of this difference, AML 2.3 was randomized 20,000 times and the standard deviation of each set of ER differences was recorded, as shown in Fig. S2. This gave a z-score of –96.8 for the ER differences in the real network. Since none of the sampled randomized networks had an ER difference width less than that of the real network, an upper limit of $5.0 \times 10^{-5}$ was placed on the p-value. The same procedure was used to find the significance in the age difference between connected genes, which resulted in a z-score of –72.0 and an upper limit of $5.0 \times 10^{-5}$ for the p-value (see Figs. S3 and S4).

**Centrality measures.** The average centrality for a given gene is defined as the mean of its degree, betweenness centrality, and PageRank. Each of these quantities was scaled such that the minimum and maximum values were 0 and 1, respectively. The inverted ER was similarly scaled to the range [0,1]. The composite score is the mean of the average centrality and inverted ER. Significance values for the number of mutations in the top 5% for all three rankings shown in Figs. 3A and 3B were computed using a hypergeometric distribution.

**Global, set, and interset efficiency.** The global efficiency (34) of a network is defined as

$$E_{global} = \frac{1}{n(n-1)} \sum_{i \neq j} \frac{1}{d_{ij}}$$

where $n$ is the number of nodes in the network, $d_{ij}$ is the unweighted distance from node $j$ to node $i$, and $0 \leq E_{global} \leq 1$ for unweighted networks. We define the *set efficiency* (SE) of a set of nodes $M$ as

$$E_M = \frac{1}{|M|(|M|-1)} \sum_{\substack{i,j \in M, \\ i \neq j}} \frac{1}{d_{ij}}$$

where $|M|$ is the number of nodes in $M$, and $0 \leq E_M \leq 1$ for unweighted networks. $E_M > E_{global}$ implies that nodes in $M$ are closer to each other than average in the network, and $E_M < E_{global}$ implies that the nodes are more dispersed than average. Note that $d_{ij}$ is calculated using the full network, so shortest paths from $j$ to $i$ may pass through nodes which are not in $M$. The SE was thus used to examine the topological distribution of ERs and ages in AML 2.3. The ERs were sorted from coldest to hottest, and the SE of the first 500 genes was computed, increasing the window size in steps of 10 genes from the beginning to the end of the ER list. The resulting curve is shown in blue in Fig. 2B. Also pictured is the mean of the controls (solid green line) plus/minus one standard deviation (dashed green lines), which were obtained by randomizing the order of the genes 100 times and computing the cumulative SE in the same manner. See Fig. S6 for the same plot using age rather than ER.

We define the *interset efficiency* (IE) from node set $J$ to node set $I$ as

$$E_{IJ} = \frac{1}{|I||J| - |I \cap J|} \sum_{\substack{i \in I, j \in J, \\ i \neq j}} \frac{1}{d_{ij}}$$

where $|I \cap J|$ is the number of nodes shared by sets $I$ and $J$, and $0 \leq E_{IJ} \leq 1$ for unweighted networks. As with the set efficiency, shortest paths may pass through nodes which are neither in $I$ nor $J$. Note that this formulation is defined when sets $I$ and $J$ have a non-empty intersection, and that the diagonal terms of the interset efficiency reduce to the set efficiency, i.e. $E_{II} = E_I$. A large $E_{IJ}$ implies that the average distance from nodes in $J$ to nodes in $I$ is small, and a small $E_{IJ}$ implies large distances. $E_{IJ}$ is asymmetric for directed networks. This measure was used in Fig. 3 to quantify the proximity of the functional groups from Table 1.

## Acknowledgements

This work was supported by NIH NCI 1R41CA174059-01 and NSF IIP 1346482. The authors thank Yunyi Kang (Sanford-Burnham-Prebys Medical Discovery Institute), Christopher Wills (UCSD), and Francesca Mulas (UCSD) for providing comments and suggestions on this manuscript.

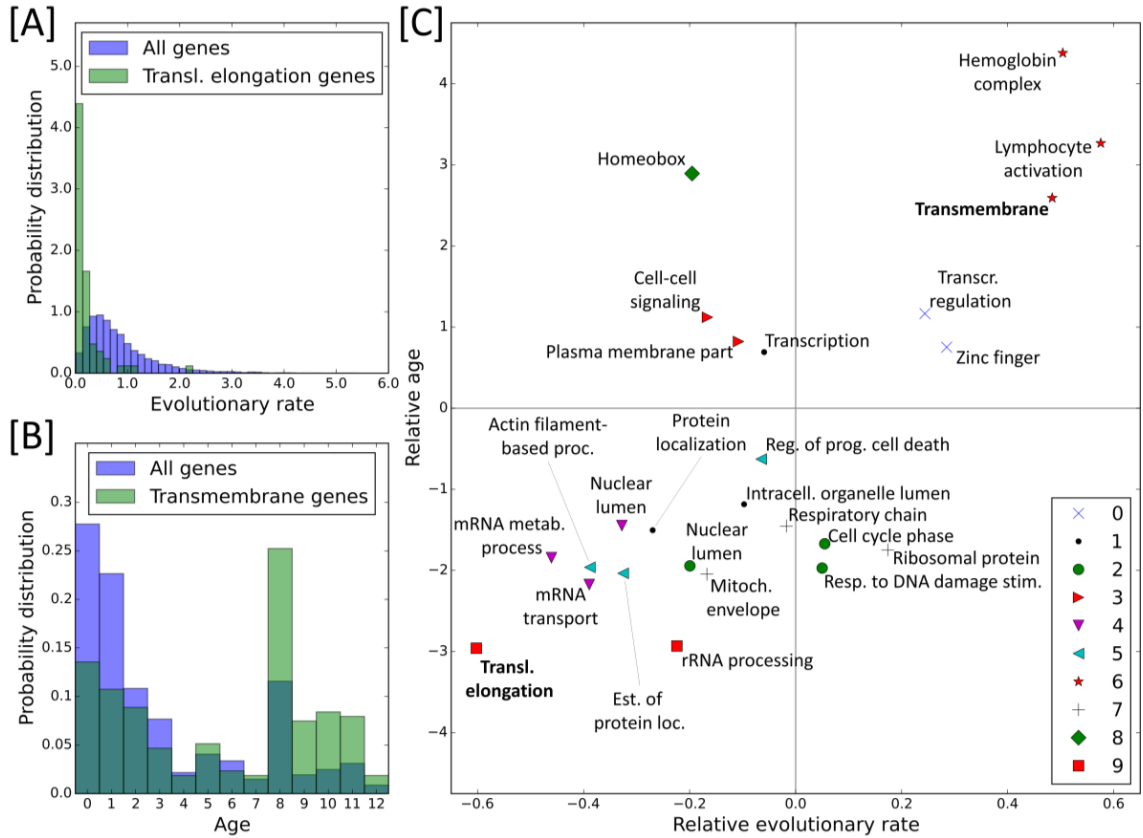

Figure 1. [A] Distribution of evolutionary rates (ERs), measured in units of the number of nonsynonymous substitutions per amino acid site per billion years, for all genes (purple) and for genes in the translational elongation functional group (green). This functional group has a very low ER compared to the background distribution. [B] Distribution of ages for all genes (purple) and genes in the transmembrane functional group (purple), where age=0 is the oldest and age=12 is the youngest. Transmembrane genes are much younger than average. [C] Summary of average ER and age for communities and functional groups in Table 1. The x-values are computed from $ER_{relative} = ER_{func.\ group\ mean} - ER_{network\ mean}$, and likewise for relative age on the y-axis. The functional groups from [A] and [B] have bold labels in [C]. Each marker type corresponds to one of communities 0 through 9. As expected, old functional groups tend to have a low average ER (i.e. are "cold"), and young functional groups tend to evolve frequently (i.e. are "hot"). Unabbreviated functional group names are listed in Table 1.

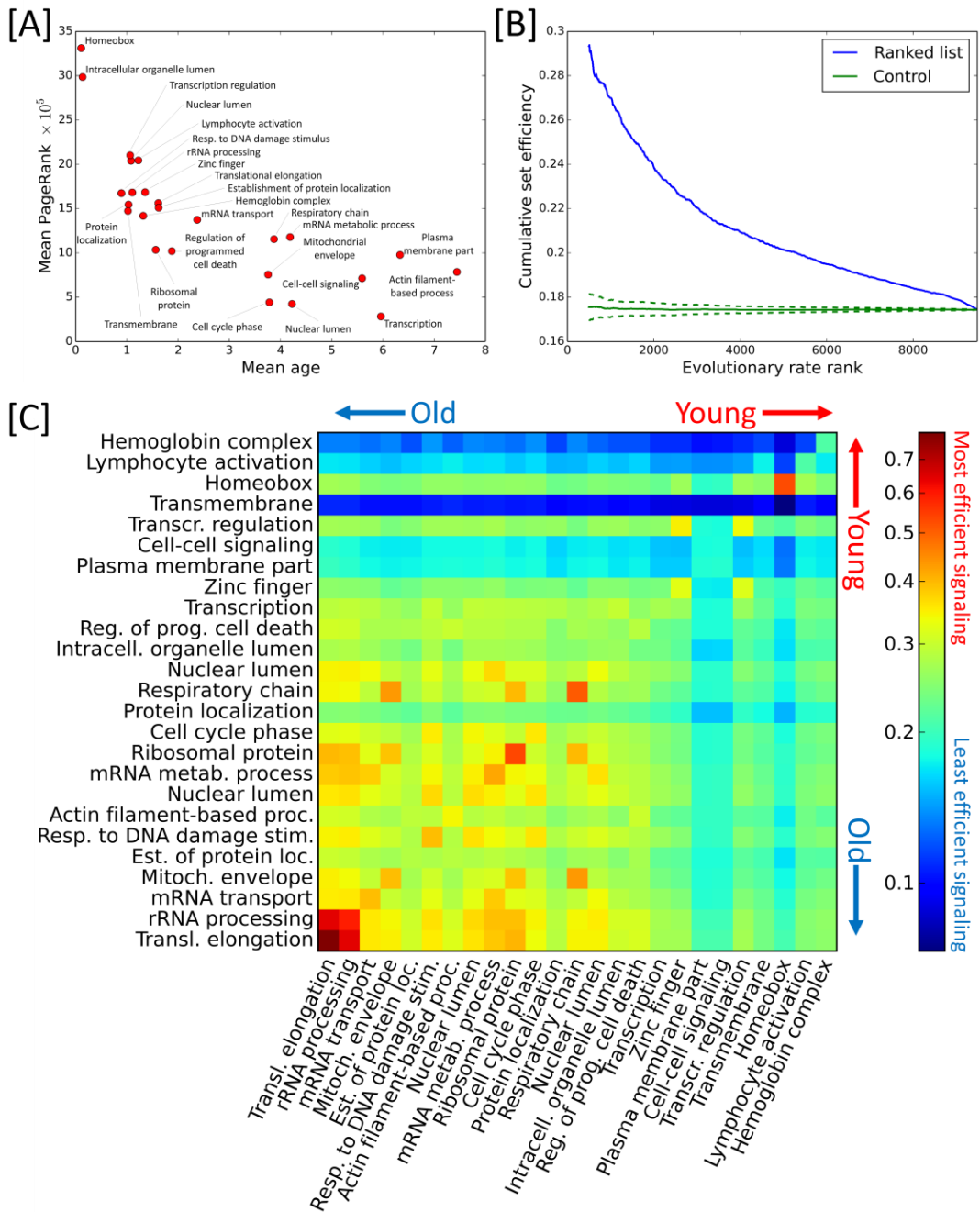

Figure 2. [A] Mean PageRank versus mean age each functional group from Table 1, where age=0 is the oldest and age=12 is the youngest. Old functional groups have high PageRank, and vice versa. Unabbreviated functional group names are listed in Table 1. [B] The cumulative set efficiency (SE) of all genes below a given evolutionary rate (ER) rank (lowest to highest ER, i.e. "coldest" to "hottest"). The SE of the 500 coldest genes is significantly higher than the control, and including hotter genes monotonically decreases the SE. This indicates that the coldest genes exchange information efficiently, while the hottest genes are more dispersed and thus communicate less efficiently than average. [C] Interset efficiency from functional group in column $j$ to functional group in row $i$. The list of functional groups was sorted by average age from oldest (transcriptional elongation) to youngest (hemoglobin complex). There is a highly efficient exchange of information between old functional groups, as indicated by the high interset efficiency values in the lower left corner. Recently developed functional groups, particularly the blood cell-specific functional groups of lymphocyte activation and hemoglobin complex, are remote from most other functional groups. Note that the above matrix is asymmetric because the network is directed, and that the colors are log-scaled.

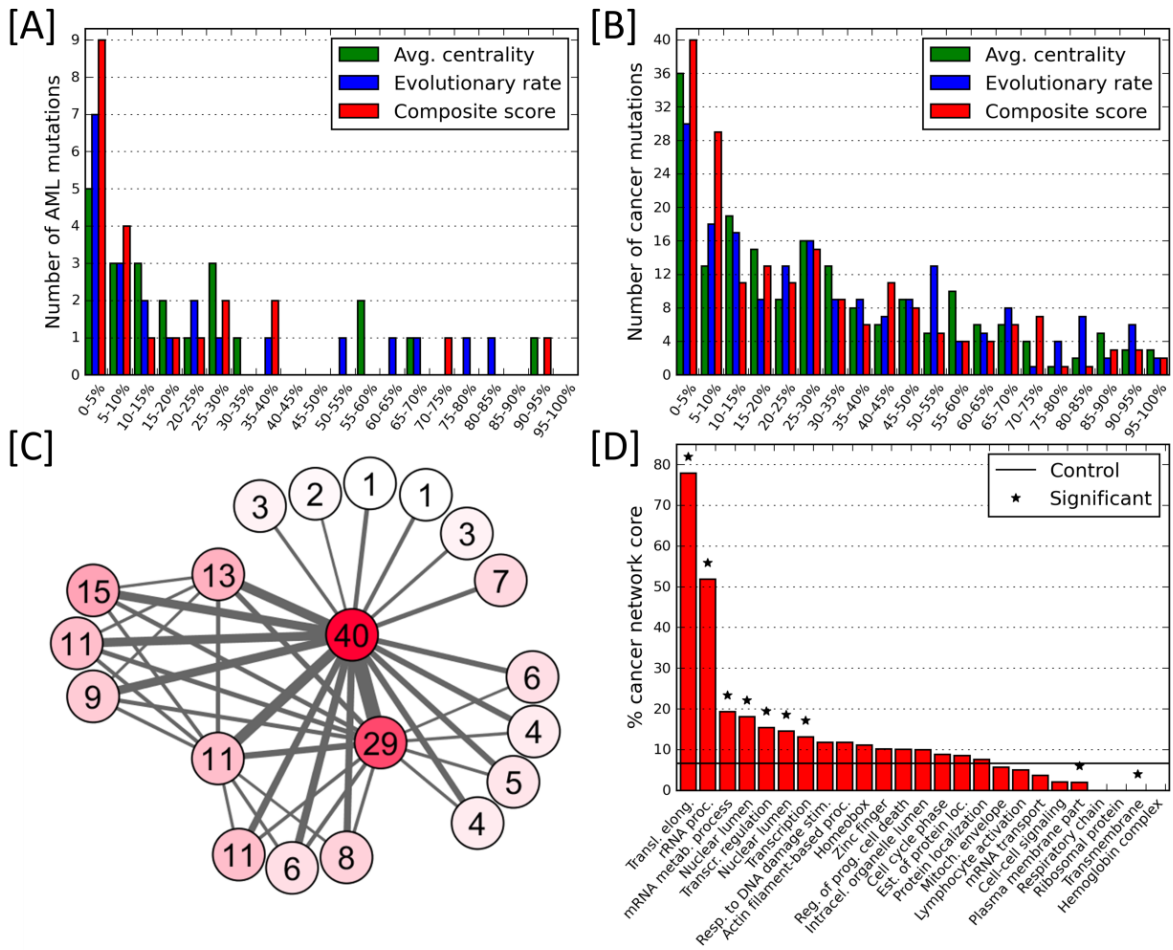

Figure 3. [A] Genes ranked by average centrality (green, highest centrality on the left), evolutionary rate (blue, lowest rate on the left) and a composite score including both quantities (red). The ranked values are divided into partitions, each including 5% of the genes. The height of each bar indicates the number of the 22 most common acute myeloid leukemia (AML) mutations present in each partition. [B] Genes ranked in the same way as in Fig. 3A, but for the 198 most common mutations across 21 different types of cancer. [C] AML network composed by nodes representing each of the 20 composite score partitions from panels A and B. Edges show the strength of connections between the genes within each partition. The numbers within the nodes show the cancer mutations, as in panel B. We define the central node, corresponding to the partition with the highest composite score, as the "cancer network core." This is the node with the highest number of mutations and connections to the other nodes. [D] The fraction of genes from the cancer network core present in the 25 functional groups from Table 1. The horizontal black line is the control, representing the expected fraction for a random set of genes of equal size. Functional groups with $p < 10^{-4}$ (computed from a hypergeometric distribution) are labelled with a star.

| Comm. index | Num. genes | Comm. ER | Diff. in mean | p-value | Comm. age | Diff. in mean | p-value | DAVID functional group | Num. genes | DAVID Benjamini | Group ER | Diff. in mean | p-value | Group age | Diff. in mean | p-value |
|---|---|---|---|---|---|---|---|---|---|---|---|---|---|---|---|---|
| 0 | 1760 | hot | 0.24 | 2.5E-81 | young | 1.01 | 6.7E-28 | Zinc finger | 275 | 2.6E-39 | hot | 0.32 | 6.0E-19 | young | 0.75 | 4.4E-05 |
| | | | | | | | | Transcription regulation | 298 | 1.9E-16 | hot | 0.28 | 4.8E-14 | young | 1.17 | 9.3E-11 |
| 1 | 1579 | average | -0.03 | 2.3E-02 | old | -0.39 | 6.8E-07 | Transcription | 304 | 4.8E-24 | average | -0.05 | 4.6E-02 | young | 0.69 | 5.6E-05 |
| | | | | | | | | Intracellular organelle lumen | 239 | 9.7E-19 | average | -0.07 | 7.1E-02 | old | -1.18 | 1.7E-08 |
| | | | | | | | | Protein localization | 145 | 2.9E-10 | cold | -0.27 | 7.3E-05 | old | -1.50 | 5.2E-08 |
| 2 | 1208 | cold | -0.10 | 2.6E-10 | old | -1.30 | 1.2E-58 | Nuclear lumen | 332 | 8.4E-107 | cold | -0.21 | 2.6E-07 | old | -1.94 | 1.6E-28 |
| | | | | | | | | Cell cycle phase | 169 | 3.6E-98 | average | 0.06 | 1.0E-01 | old | -1.67 | 5.4E-07 |
| | | | | | | | | Response to DNA damage stimulus | 135 | 3.5E-58 | average | 0.03 | 3.1E-01 | old | -1.97 | 3.2E-07 |
| 3 | 1055 | average | 0.01 | 2.4E-01 | young | 1.41 | 5.4E-53 | Cell-cell signaling | 146 | 3.6E-39 | cold | -0.16 | 7.0E-04 | young | 1.12 | 1.3E-05 |
| | | | | | | | | Plasma membrane part | 359 | 1.1E-67 | cold | -0.11 | 2.5E-04 | young | 0.82 | 4.1E-06 |
| 4 | 867 | cold | -0.27 | 3.7E-55 | old | -0.94 | 1.7E-16 | mRNA metabolic process | 124 | 9.5E-66 | cold | -0.46 | 2.7E-25 | old | -1.84 | 2.7E-06 |
| | | | | | | | | Nuclear lumen | 212 | 9.3E-60 | cold | -0.34 | 5.3E-24 | old | -1.45 | 1.2E-09 |
| | | | | | | | | mRNA transport | 27 | 1.2E-11 | average | -0.36 | 5.8E-03 | average | -2.17 | 7.2E-03 |
| 5 | 780 | cold | -0.11 | 1.1E-05 | old | -0.91 | 5.1E-10 | Establishment of protein | 105 | 1.1E-19 | cold | -0.33 | 5.2E-09 | old | -2.03 | 1.6E-12 |
| | | | | | | | | Actin filament-based process | 51 | 1.9E-16 | cold | -0.41 | 9.0E-06 | old | -1.96 | 9.0E-06 |
| | | | | | | | | Regulation of programmed cell | 79 | 2.6E-07 | average | -0.03 | 4.2E-01 | average | -0.63 | 3.0E-02 |
| 6 | 748 | hot | 0.18 | 3.5E-22 | young | 1.34 | 2.8E-29 | Lymphocyte activation | 40 | 1.2E-11 | hot | 0.68 | 1.2E-10 | young | 3.27 | 6.2E-11 |
| | | | | | | | | Hemoglobin complex | 13 | 1.3E-12 | average | 0.54 | 5.1E-03 | young | 4.43 | 1.1E-04 |
| | | | | | | | | Transmembrane | 252 | 1.8E-05 | hot | 0.51 | 2.0E-34 | young | 2.55 | 4.5E-28 |
| 7 | 417 | average | -0.08 | 1.2E-02 | old | -1.54 | 8.3E-22 | Mitochondrial envelope | 103 | 1.0E-68 | average | -0.14 | 2.0E-02 | old | -1.98 | 4.5E-10 |
| | | | | | | | | Respiratory chain | 44 | 4.9E-47 | average | 0.03 | 3.4E-01 | average | -1.57 | 9.3E-05 |
| | | | | | | | | Ribosomal protein | 62 | 2.4E-53 | average | 0.18 | 2.3E-02 | old | -1.73 | 3.2E-05 |
| 8 | 296 | hot | 0.16 | 1.5E-05 | young | 1.49 | 4.8E-14 | Homeobox | 27 | 2.2E-12 | average | -0.22 | 7.3E-02 | young | 3.01 | 3.7E-07 |
| 9 | 270 | cold | -0.28 | 5.7E-14 | old | -1.96 | 2.9E-20 | Translational elongation | 77 | 1.7E-114 | cold | -0.60 | 6.0E-18 | old | -3.04 | 1.7E-13 |
| | | | | | | | | rRNA processing | 27 | 1.1E-22 | average | -0.24 | 3.2E-02 | old | -2.93 | 1.7E-05 |

Table 1. Community and functional group structures present in our acute myeloid leukemia network, AML 2.3, with corresponding evolutionary rate (ER) and age properties. Gene ERs take real values from 0 (most conserved) to approximately 6.9 (most variable), and ages take integer values from 0 (oldest) to 12 (youngest). The table is organized as follows: "Comm. Index" is the index of the ten largest communities; "Num. genes" is the number of genes in the community; "Comm. ER" indicates whether the community is significantly hotter (i.e. has a higher evolutionary rate) or colder (i.e. has a lower evolutionary rate) than the mean of 300 equally-sized sets of randomly selected genes, with a significance threshold of $p = 10^{-3}$; "Diff. in mean" is the difference between the mean ER of the community and the mean ER of the 300 randomly selected sets; "p-value" is the significance of the difference; "Comm. age", "Diff. in mean", and "p-value" are the same as previously stated, but for age rather than ER; "DAVID functional group" is the name of the functional groups that DAVID identified as enriched in each community; "Num. genes" is the number of genes in the functional group; "DAVID Benjamini" is the significance of the enrichment of the functional group, as reported by DAVID; and the remaining functional group columns are computed in the same manner as the community columns.